Highlights

**SBcoyote: An Extensible Python-Based Reaction Editor and Viewer**

Jin Xu,Gary Geng,Nhan D. Nguyen,Carmen Perena-Cortes,Claire Samuels,Herbert M. Sauro

- Research highlights item 1: SBcoyote is a fully Python-based biochemical reaction editor and viewer.
- Research highlights item 2: SBcoyote plugins support SBML level 3 with layout and render.
- Research highlights item 3: SBcoyote is extensible by a third party via its Python plugin API.

# SBcoyote: An Extensible Python-Based Reaction Editor and Viewer

Jin Xu[a,1], Gary Geng[b,1], Nhan D. Nguyen[c], Carmen Perena-Cortes[d], Claire Samuels[d] and Herbert M. Sauro[a,*]

[a]*Department of Bioengineering, University of Washington, Seattle 98195, WA, USA*
[b]*Department of Computer Science, University of Washington, Seattle 98195, WA, USA*
[c]*Department of Chemistry and Biochemistry, Augustana University, Sioux Falls, 57197, SD, USA*
[d]*Department of Mathematics, University of Washington, Seattle 98195, WA, USA*

ABSTRACT

SBcoyote is an open-source cross-platform biochemical reaction viewer and editor released under the liberal MIT license. It is written in Python and uses wxPython to implement the GUI and the drawing canvas. It supports the visualization and editing of compartments, species, and reactions. It includes many options to stylize each of these components. For instance, species can be in different colors and shapes. Other core features include the ability to create alias nodes, alignment of groups of nodes, network zooming, as well as an interactive bird-eye view of the network to allow easy navigation on large networks. A unique feature of the tool is the extensive Python plugin API, where third-party developers can include new functionality. To assist third-party plugin developers, we provide a variety of sample plugins, including, random network generation, a simple auto layout tool, export to Antimony, export SBML, import SBML, etc. Of particular interest are the export and import SBML plugins since these support the SBML level 3 layout and render standard, which is exchangeable with other software packages. Plugins are stored in a GitHub repository, and an included plugin manager can retrieve and install new plugins from the repository on demand. Plugins have version metadata associated with them to make it install plugin updates. Availability: https://github.com/sys-bio/SBcoyote.

## 1. Introduction

Python has become a very popular open-source language for scientific computing and data science as it is easy to learn and use. The systems biology community has shown interest in using Python through the development of a variety of simulation tools. For example, Tellurium (Choi et al., 2018; Medley et al., 2018) provides an extensible environment for modeling; PySCeS (Olivier et al., 2005) focuses on simulation via differential equations, structural analysis, and metabolic control analysis; SloppyCell (Myers et al., 2007) focuses on model fitting and calculating the resulting uncertainties; pySB (Lopez et al., 2013) focuses on rule-based reaction models; and COBRApy (Lerman et al., 2013) focuses on constraint-based modeling. However, other than SBMLDiagrams (Xu et al., 2022), there are few tools written in Python for visualizing reaction networks. SBMLDiagrams is a command line tool for displaying biochemical networks and uses the SBML layout and render extensions to manage the visual information. In this article, we wish to describe a new Python-based tool called SBcoyote. This is an extensible, cross-platform SBML-compatible reaction viewer and editor. It supports the visualization of compartments, species, and reactions and includes many options to stylize each of these components. SBcoyote's primary purpose is to enable users to add layout and render information to an SBML model or to edit an existing SBML model with layout and render. SBcoyote is extensible through the use of user-written Python plugins. This makes it easy for third-party users to add new functionality to SBcoyote. Certain aspects of SBML, such as annotations, events, and the various SBML rules, cannot be currently edited using SBcoyote. However, it wouldn't be difficult to create plugins that could support these features.

In addition to the core features, we provide a set of sample plugins. These include plugins to generate random networks, auto layout of networks, export models in the form of Antimony (Smith et al., 2009), export and import of SBML (Hucka et al., 2003), and a variety of other useful functional plugins such as aligning nodes in a circular pattern and carrying out the basic structural analysis of a network (Vallabhajosyula et al., 2006). Of particular interest are the SBML export and import plugins that support the SBML level 3 layout and render standard (Deckard et al., 2007). A critical issue in systems biology is ensuring the exchange and reproducibility of models (Choi et al., 2018; Porubsky et al., 2020). Over the past few years, the community has developed a variety of standards to accurately capture models and simulation experiments. One popular standard is the Systems Biology Markup Language (SBML) (Hucka et al., 2003). SBML is used primarily to allow the exchange of biochemical models between different software tools. SBML level 3 also includes an extension that allows diagrammatic information to be stored in the form of layout and render information. The layout component describes the positions and sizes of different graphical objects, including compartments, species, and reactions. The render component describes color and shape information. There are currently three tools (not including SBcoyote) that support the layout and render information in SBML level 3 well. These include SBMLDiagrams (Xu et al., 2022), COPASI (Hoops et al., 2007) and MINERVA (Gawron et al., 2016).

[*]Corresponding author
hsauro@uw.edu (H.M. Sauro)
ORCID(s): 0000-0001-6738-9979 (J. Xu); 0000-0002-3659-6817 (H.M. Sauro)
[1]Jin Xu and Gary Geng contributed equally.

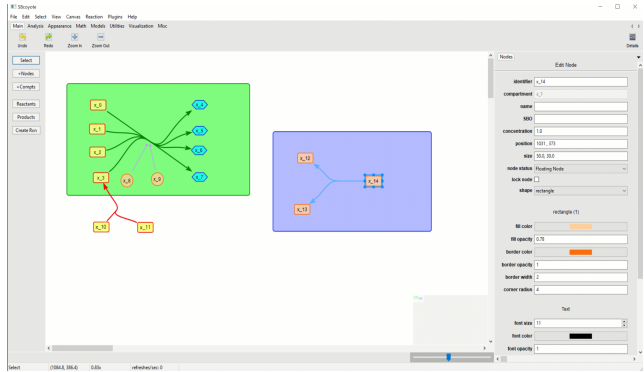

**Figure 1: SBcoyote GUI example.** There are two compartments with different fill and border colors. Nodes can be in different shapes and colors. The reactions are in bezier curves with different colors. SBcoyote supports modifiers too. The interactive bird-eye view of the network is at the down-right corner. Users can modify node features in the right-hand form after selecting a certain node.

A unique feature of SBcoyote is the extensible plugin system. Third-party developers can include new functionality via the SBcoyote plugin API. Most tools in systems biology are not extensible. For example, network visualization tools such as Escher (King et al., 2015a), Newt (Balci et al., 2020), and VCell (Schaff et al., 1997) are not extensible. MINERVA (Gawron et al., 2016), Cytoscape (Shannon et al., 2003), PathwayDesigner (Bedaso et al., 2018) (https://www.pathwaydesigner.org/), and CellDesigner (Funahashi et al., 2006) do support the addition of plugins, but users require high-level programming skills to develop them since they need to be competent in Java, or Object Pascal. SBcoyote, however, uses Python as the language for writing plugins, and Python is a relatively easy language to learn.

## 2. Core Features

SBcoyote is a fully Python-based cross-platform, biochemical reaction viewer and editor. It uses wxPython to implement the GUI and the drawing canvas. Figure 1 shows a screenshot of the SBcoyote GUI. Users can add, arrange and modify compartments, species, and reactions. It can also display regulatory edges, such as feedback loops between species and reactions. It includes many options to stylize each of these components. These include:

- Compartments can be visualized in different positions, sizes, and colors.

- Support is provided for floating and boundary species. Species can be visualized in different positions, sizes, colors, and shapes as well as grouped for alignment purposes.

- Species nodes can be duplicated to form alias species to help visualize networks where a given species is used in many reactions.

- SBcoyote supports complex species that use some aspects from the Systems Biology Graphical Notation (SBGN) (Novère et al., 2009).

- Reactions can be drawn either using bezier curves or straight lines, and reactions can be displayed in different colors, thicknesses etc.

- SBcoyote supports zooming as an interactive bird-eye view of a network to allow easy navigation of large networks.

- SBcoyote can export and import models using SBML and JSON formats.

Figure 2A and B illustrate two visualization examples of the metabolic network from Jana Wolf's work (Wolf et al., 2001). Figure 2C shows a visualization example of the large-scale Escherichia coli (ecoli) core metabolism network (King et al., 2015b; Orth et al., 2010).

## 3. Plugins

SBcoyote is extensible due to its Python plugin API, where third-party developers can include new functionality. A range of sample plugins is provided with the distribution. A plugin manager is provided that gives version metadata for installing or updating plugins. Plugins are stored in a GitHub repository (https://github.com/sys-bio/SBcoyote-plugins), and an included plugin manager can retrieve and install new plugins from the repository on demand. Plugins have version metadata associated with them to make it easy to install new versions.

### 3.1. Supplied Plugin Examples

Plugins can be accessed by either selecting the plugins button on the Navigation Bar or by selecting the Applications Menu. In this section, we will describe some of the sample plugins supplied in the distribution.

#### 3.1.1. Random Network

This plugin creates random reaction networks, shown in Figure 3A. Users can specify the size of the networks by choosing the number of species and reactions. In addition, the probability of each type of mass-action reaction, i.e. UniUni, BiUni, UniBi, BiBi, can be adjusted as long as the sum of the four numbers is one. It also allows users to assign a random seed to generate the same random network multiple times. This is useful for testing purposes. Orphan nodes that are not connected to any reactions are automatically removed.

#### 3.1.2. Auto Layout

The auto layout plugin will automatically layout a network. This is useful for laying out random networks or loading SBML models which do not have information on the layout of the model. The plugin uses the spring algorithm (Fruchterman and Reingold, 1991) from NetworkX (Hagberg et al., 2008), see Figure 3B. Users can change the optimal distance between nodes, the scale of the layout and also choose to arrange the reaction centroids or not.

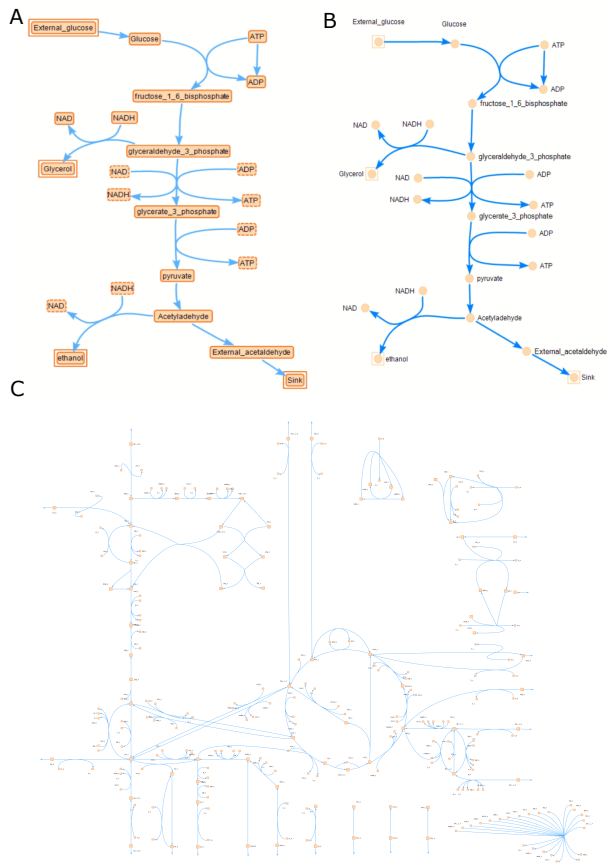

**Figure 2: Some visualization examples by SBcoyote.** A. Using SBcoyote to visualize a model of glycolysis (Wolf et al., 2001). Alias nodes are indicated with dashed border lines, and boundary nodes have an extra square outside compared with floating nodes. B. Another visualization for the model of glycolysis (Wolf et al., 2001) but with nodes in circles and texts outside the nodes. C. Using SBcoyote to visualize the large-scale Escherichia coli (ecoli) core metabolism network (King et al., 2015b; Orth et al., 2010). The figure with a larger size is also available on GitHub (https://github.com/sys-bio/SBcoyote).

### 3.1.3. Export Antimony

The export Antimony plugin allows a network to be exported as an Antimony string (Smith et al., 2009). Users can also copy the Antimony string to the clipboard or save it to a file. Antimony can be loaded into other tools such as Tellurium (Choi et al., 2018; Medley et al., 2018).

### 3.1.4. Export SBML

The export SBML plugin allows users to export the network to SBML (Hucka et al., 2003). Users can copy the SBML to the clipboard or save it to a file. The SBML files are stored using SBML level 3 and include both layout and render information as part of the SBML. The exported SBML files from SBcoyote can be visualized by import SBML plugin (Figure 3C) or other software tools such as COPASI (Figure 3D).

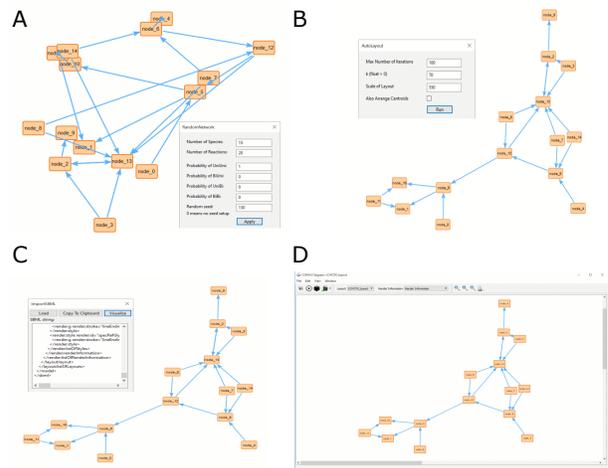

**Figure 3: Plugin samples.** A. A reaction network is generated by the random network plugin. B. The auto layout plugin makes the reaction network easier to examine. C. The import SBML plugin can import an SBML file back to SBcoyote after generating the SBML file by the export SBML plugin. D. The visualization by COPASI GUI of the SBML exported from SBcoyote shows the reproducibility and exchangeability of the network layout.

### 3.1.5. Import SBML

The import SBML plugin lets a user import SBML. Once loaded, the software will check for the SBML layout information and, if found, will use it to visualize the network. If no layout information is found, the species in the network are randomly positioned on the canvas. A user can now use the auto layout plugin or manually rearrange the network. When exported to SBML, the new layout is stored in the SBML. See Figure 3C.

### 3.1.6. Align Circle

The align circle plugin is a simple plugin that will align all the selected nodes in a circular pattern with either the default radius or a desired radius.

### 3.1.7. Structural Analysis

The structural analysis plugin can be used to visualize the stoichiometry matrix and conservation matrix of the reaction network on the canvas. It can compute the stoichiometry matrix and the conservation matrix of the network, and can also highlight a given conserved moiety using a selected color and un-highlight the nodes afterward. See Figure 4.

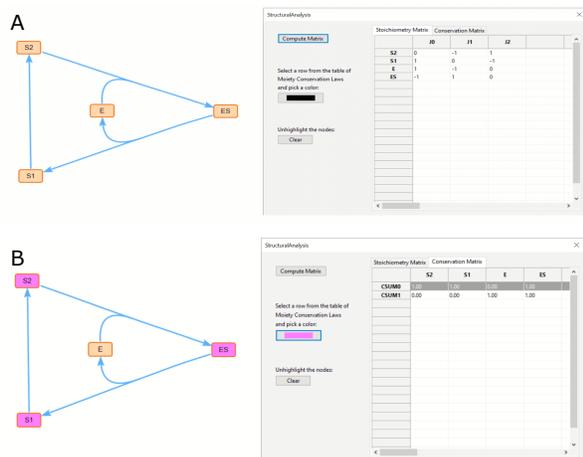

**Figure 4: Structural analysis plugin.** A. Compute and show the stoichiometry matrix. B. Compute the conservation matrix and highlight the nodes of a conserved moiety.

### 3.2. SBML Compliance

The export and import SBML plugin supports SBML level 3 but, most importantly, the layout and render extension. Layout describes the position and size of different graphical objects, while render describes the color, shape, and line styles. Figure 3C shows an example of a SBML model with layout and render being used to correctly display the network. SBML layout and render is compatible with other tools such as SBMLDiagrams (Xu et al., 2022), CO-PASI (Hoops et al., 2007) and MINERVA (Gawron et al., 2016). Figure 3D shows a visualization by COPASI GUI of the SBML file exported from SBcoyote. The export and import of SBML are also available under the File of the Applications Menu with a single mouse click.

### 3.3. Developing Plugins

As a third party, creating new functionality via the SBcoyote Python plugin API is straightforward. There are two plugin types. Users can develop either a Command-Plugin or a WindowedPlugin. CommandPlugins are those with one single action and do not require a dialog display. WindowedPlugin allows for a more complex UI by spawning a window. There are seven plugin categories, i.e., ANALYSIS, APPEARANCE, MATH, MODELS, UTILITIES, VISUALIZATION, MISC. Developers can assign the plugin category in the PluginMetadata object. All the plugin categories are availble in the Applications Menu. The API is well-documented at https://sys-bio.github.io/SBcoyote/api.html. For example, in the random network plugin, adding a random node to the canvas can be achieved using the api.add_node() function with a random position assigned to the parameter of position.

### 4. Conclusion and Discussion

SBcoyote was an experiment to see if we could write a biochemical network editor in Python. On the whole, we believe we have succeeded. The application may suffer some slight sluggishness, though it is by no means severe; for example, editing large networks (such as the ecoli model shown in Figure 2C, which has 95 reactions and 72 species), is possible. We hope to move eventually to Python 3.11, which is reported to be 1.25x faster than previous Python versions. Currently, some dependencies are not yet 3.11 compatible.

SBcoyote provides the systems biology community with an extensible Python-based reaction network editor and viewer. SBcoyote is cross-platform and available under the liberal MIT license. The tool can visualize and modify reaction networks with its core features. A unique feature of the tool is the extensive Python plugin API, where third-party developers can include their new functionality. The SBML export and import plugins support the standard of SBML level 3 layout and render, allowing models to be exchangeable with other visualization tools through SBML. See Table 1 for the summary of feature comparisons among different reaction network viewers and editors from the perspectives of SBML level 3 with layout and render, and plugin APIs.

There are limitations to SBcoyote due to the use of wxPython and Python. Currently, loading large networks is slow. For example, the ecoli model in Figure 2C takes 2 minutes to load on an Intel i9-9900XCPU 3.50 GHz, but we are investigating ways to speed this up by optimizing the Python code.

**Table 1**
Comparison of different reaction network viewers and editors from the perspectives of SBML level 3 with layout and render and plugin APIs.

| Tools | SBML3 layout and render | Plugin APIs (Language) |
|---|---|---|
| **SBcoyote** | Yes | Yes (Python) |
| **SBMLDiagrams** | Yes | No |
| **COPASI** | Yes | No |
| **MINERVA** | Yes | Yes (Java) |
| **Escher** | No | No |
| **Newt** | Yes | No |
| **VCell** | No | No |
| **Cytoscape** | No | Yes (Java) |
| **CellDesigner** | No | Yes (Java) |

### Availability

SBcoyote is publicly available and under the liberal MIT open-source license. The source code has been deposited at GitHub (https://github.com/sys-bio/SBcoyote). Sample plugins are also available on GitHub (https://github.com/sys-bio/SBcoyote-plugins). The package is fully documented at (https://sys-bio.github.io/SBcoyote).

# Author Contributions

J.X. and G.G. contributed equally to this work. J.X. was the main developer of the plugins, added the compliance of SBML level 3 layout and render, helped with some core features and documentation, and wrote the manuscript. G.G. was the primary developer of the core features with Docstrings written for the documentation. N.D.N. helped develop the node shapes. C.P.-C. was the main developer of the documentation and the auto layout plugin. C.S. developed some plugins, created the plugin version metadata, and helped with the documentation. H.M.S. conceived the idea, was responsible for the project administration and funding acquisition, helped with coding some plugins and core features, and wrote the manuscript.

# Acknowledgement

This work was supported by the National Institutes of Health [U24EB028887]. The content is solely the responsibility of the authors and does not necessarily represent the official views of the National Institutes of Health or the University of Washington. J.X. thanks Frank T. Bergmann for his assistance in using python-libSBML. G.G. thanks Renjie Zhou, who initiated the iodine.py before his adaptation. We also thank Evan Yip, who implemented the align circle plugin.

# Declares of Interest

None.